\documentclass[conference]{IEEEtran}
\IEEEoverridecommandlockouts
\usepackage{cite}
\usepackage{amsmath,amssymb,amsfonts}
\usepackage{algorithmic}
\usepackage{graphicx}
\usepackage{hyperref}
\usepackage{textcomp}
\usepackage{xcolor}
\usepackage{listings}
\usepackage{multirow}

\begin{document}

\title{SECOMlint: A linter for Security Commit Messages}

\makeatletter
\newcommand{\linebreakand}{%
  \end{@IEEEauthorhalign}
  \hfill\mbox{}\par
  \mbox{}\hfill\begin{@IEEEauthorhalign}
}
\makeatother

\author{\IEEEauthorblockN{Sofia Reis}
\IEEEauthorblockA{\textit{INESC-ID \& IST, U. Lisbon} \\
Lisbon, Portugal \\
sofia.o.reis@tecnico.ulisboa.pt}
\and
\IEEEauthorblockN{Corina P\u{a}s\u{a}reanu}
\IEEEauthorblockA{\textit{Carnegie Mellon University} \\
USA \\
pcorina@cmu.edu}
\linebreakand
\IEEEauthorblockN{Rui Abreu}
\IEEEauthorblockA{\textit{INESC-ID \& FEUP, U. Porto} \\
Porto, Portugal \\
rui@computer.org}
\and
\IEEEauthorblockN{Hakan Erdogmus}
\IEEEauthorblockA{\textit{Carnegie Mellon University} \\
USA \\
hakan.erdogmus@west.cmu.edu}
}

\maketitle

\begin{abstract}
Transparent and efficient vulnerability and patch disclosure 
are still a challenge in the security community, essentially 
because of the poor-quality documentation stemming from the lack of standards.
SECOM is a recently-proposed standard convention for security commit messages 
that enables the writing of well-structured 
and complete commit messages for security patches. The
convention prescribes different bits of security-related 
information essential for a better understanding of vulnerabilities
by humans and tools.
SECOMlint is an automated and configurable solution to help 
security and maintenance teams infer compliance against the SECOM
standard when submitting patches to security vulnerabilities 
in their source version control systems. The tool leverages the natural 
language processing technique Named-Entity Recognition (NER) to 
extract security-related information from commit messages and 
uses it to match the compliance standards designed. We provide a 
demonstration of SECOMlint at \url{https://youtu.be/-1hzpMN_uFI}; and documentation and 
its source code at \href{https://tqrg.github.io/secomlint/}{https://tqrg.github.io/secomlint/}.
\end{abstract}


\section{Introduction}
Only $9\%$ of known security vulnerabilities reported 
on websites such as the Open-Source Vulnerability (OSV)\footnote{\url{https://osv.dev/}} database 
and the National Vulnerability database (NVD)\footnote{\url{https://www.nist.gov/programs-projects/national-vulnerability-database-nvd}} reference 
the fixes, i.e., the code sample (or diff) that patched
the problem. Maintenance and security teams revisit many of these vulnerabilities 
in yearly security reports or even when new exploits are found (in 
case of an incomplete fix). In addition, researchers often use the history 
of vulnerabilities to learn more about their scope~\cite{10.1145/3133956.3134072} and as a baseline to 
validate new approaches in the security field~\cite{10.1145/3533767.3534380,maintainable-sec}.

Transparent and efficient vulnerability and patch 
disclosure is still a challenge in the security
community, essentially because of poor 
quality documentation~\cite{Tian_2022} which is the result
of the lack of standards and poor (or non) application of existent 
best practices. In the security field, many times, the 
documentation is poor on purpose to avoid potential 
threats that may arise from providing detailed information on 
vulnerabilities and respective patches. However, 
this violates \textit{transparent disclosure} and encumbers 
the life of many maintenance and security teams.
SECOM is a recently-proposed standard convention for security commit messages 
that enables the writing of well structured 
and complete commit messages for security patches~\cite{9796324}. 
The convention includes security-related 
information that is crucial for a better understanding of 
vulnerabilities and respective patches for both humans 
and tools. 

\begin{figure}[t!]
    \centering
    \includegraphics[width=\linewidth]{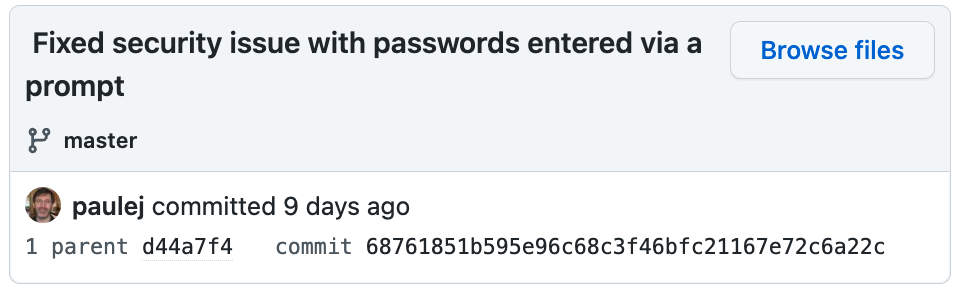}
    \caption{Example of the commit message used to patch the CVE-2022-35928 (screenshot taken on August 10th)}\label{fig:message}
    \vspace{-1.5em}
\end{figure}

\begin{figure*}[t!]
    \centering
    \includegraphics[width=\linewidth]{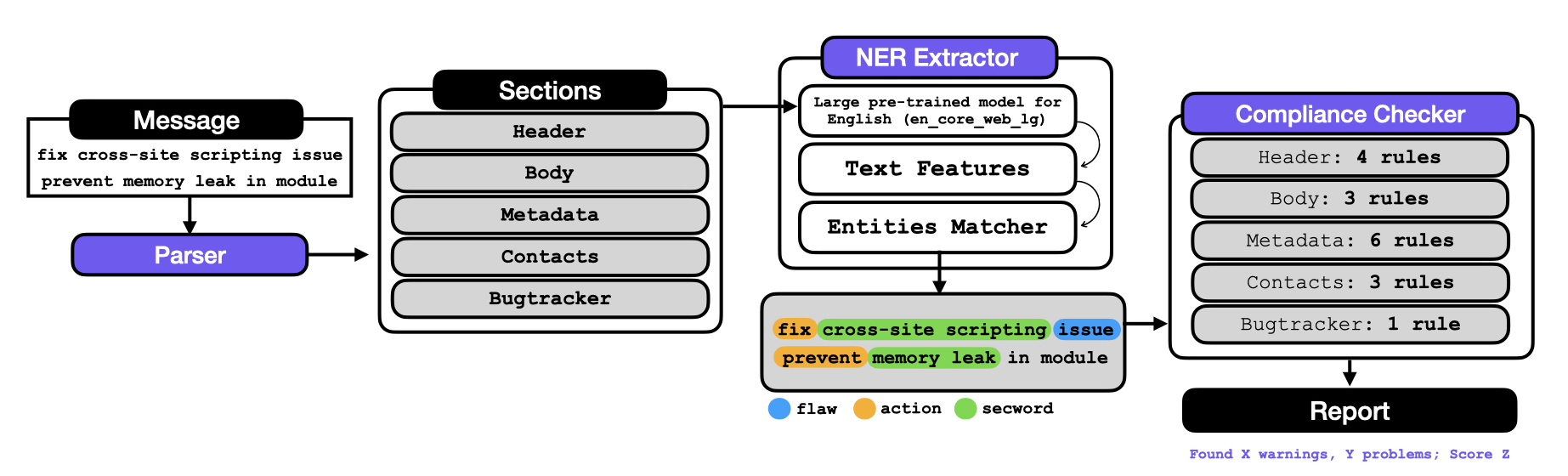}
    \caption{Execution steps and different components of \texttt{SECOMlint}.}\label{fig:components}
\end{figure*}

\textbf{Motivation.} Known vulnerabilities are
reported on websites such as the National Vulnerability Database 
(NVD) and the Open-Source Vulnerability (OSV) Database. 
One example of a known vulnerability is CVE-2022-35928\footnote{The 
CVE-2022-35928 vulnerability report is available at 
\url{https://nvd.nist.gov/vuln/detail/CVE-2022-35928}}. 
A security commit is any commit involved in the patch of a known 
security vulnerability, such as CVE-2022-35928. For instance, if you look at the section ``References to Advisories, 
Solutions, and Tools'' of the CVE-2022-35928 report, you will 
find a GitHub reference to a 
GitHub commit (\texttt{6876185}) which is the commit 
responsible for the patch. 
A security commit message is typically used to document 
the security commit used to patch the vulnerability. 
Figure~\ref{fig:message} shows the commit message used
to document the patch for CVE-2022-35928. This commit message only 
provides a very brief description of the patch. It does not describe 
the vulnerability, its importance, and the patch in detail. It also does
not provide the CVE identifier, the type of weakness, the severity of the vulnerability,
who reviewed and authored the patch, and other bits of information that are crucial 
for understanding and maintaining the vulnerability and the patch.

\textbf{Security and Maintenance Teams.} Providing 
well-structured documentation and context on the patches
of these vulnerabilities in security commit messages 
is highly important for maintenance teams when 
vulnerabilities are exploited again or revisited in 
yearly reports. 

\textbf{SECOMlint} is a linter or compliance checker that automatizes the 
verification of security commit messages against
the SECOM convention.
The linter applies rules on different text features
and calculates the degree of compliance of a message with
the standard. An output report is provided 
to assist security engineers to produce
better security commit messages.

This paper describes \texttt{SECOMlint} and 
how one can use it to produce better, more informative, security commit messages.

\section{SECOM Convention}

SECOM is a convention for security commit 
messages~\cite{9796324}. The convention was created 
based on a well-known group of 
sources~\cite{convcom, atomic, linus, goodcommit} 
on writing good commit messages to facilitate 
the adoption in practice. The structure and set of fields 
included in the convention were inferred 
from {\em 1)} an empirical analysis of security-related 
commit messages collected from vulnerability
databases such as NVD and OSV and 
{\em 2)} feedback collected alongside
Open Source Security Foundation (OpenSSF).
SECOM was considered one of the best 
practices for bulk generation of pull requests 
to scale vulnerability patching~\cite{blackhat}.

The convention (Figure~\ref{lst:SECOM}) consists of \textbf{five} main 
sections: \textbf{header}, includes the type \texttt{vuln-fix}, 
a simple description of the vulnerability and its identifier 
(when available); \textbf{body}, describes the vulnerability 
(what), its impact (why) and the patch to fix the vulnerability 
(how); \textbf{metadata}, such as type of weakness (CWE-ID), 
severity, CVSS, detection methods, report link, and version 
of the software where the vulnerability was introduced; 
\textbf{contacts}, the names and e-mail contacts of 
the \textbf{reporters} and \textbf{reviewers}; and, finally,
\textbf{references} to bug trackers. The different
sections should be separated with a new line.

\begin{lstlisting}[caption={SECOM Convention},label={lst:SECOM},frame=tlrb]
<type>: <header/subject> (<Vuln-ID>)

<body>
# (what) describe the vulnerability
# (why) describe its impact
# (how) describe the fix

Weakness: <Weakness Name/CWE-ID>
Severity: <Low, Medium, High, Critical>
CVSS: <Severity numerical representation>
Detection: <Method, Tool>
Report: <Report Link>
Introduced in: <Commit Hash>

Reported-by: <Name> (<E-mail>)
Signed-off-by: <Name> (<E-mail>)

Bug-tracker: <Bug-tracker Link>
OR
Resolves: <Issue/PR No.>
See also: <Issue/PR No.>
\end{lstlisting}

\section{Checking Security Commit Message Compliance with SECOMlint}

\texttt{SECOMlint} is written in Python and requires at least 
Python 3.8 to run. It can be run as a standalone command-line 
application, and it is released under the MIT open-source license.

\subsection{SECOMlint's Components}

Figure~\ref{fig:components} shows the different components of 
\texttt{SECOMlint} and the information flow for one run of it.
The tool takes as input a commit message which is parsed 
and divided into different sections: header, body, metadata,
contact and bug-tracker references.

For each of those sections, we apply our Named Entity Recognition (NER) 
extractor. NER (or entity chunking) is a technique used in the natural 
language processing field to identify and extract key information, 
also known as entities, in the text. Entities may be organizations, 
people's names, company names, and more. Spacy\footnote{https://spacy.io} provides
language processing pipelines that take the text and perform 
several different pre-processing steps such as \textit{tokenization} (segment 
text into tokens), \textit{tagger} (assign part-of-speech tags), 
\textit{parser} (assign dependency labels between tokens), and \textit{NER} (detect 
and label named entities). The different steps can be enabled, disabled, or 
replaced by new rule-based models, i.e., steps such as NER can 
be replaced by new rule-based models to extract new types of 
entities. From a large empirical analysis of security commit messages,
we noticed some groups of keywords that could represent different 
entities, such as vulnerability IDs, weakness IDs, severity,
security-related words, and more. These groups of keywords or 
patterns can be translated into rules and used by the NER step 
in the spacy pipeline to extract the set of entities designed.
Table~\ref{tab:entities} shows the different types of entities 
we end up designing. Some of the entities are not security 
specific but important since they are patterns that come from 
applying generic commit messages best practices, e.g., references to 
issues (\#NUMBER).

\begin{table}[t!]
    \centering
        \caption{Entities description and rationale.} 
    \begin{tabular}{| p{1.5cm} | p{6.5cm} | }
    \hline
        \textbf{entity} & \textbf{rationale}\\\hline
        ACTION & A commit usually implies an action: adding some features, 
        fixing a problem, refactoring code, and more.\\\hline
        FLAW & Fixing a security vulnerability usually implies fixing a flaw (e.g., problem,
        defect, issue, weakness, flaw, fault, bug, error, etc.). 
        \\\hline
        VUNLID & Known vulnerability ids: CVE, GHSA, OSV, etc.\\\hline
        CWEID & Vulnerability type (CWE-ID/Weakness name).\\\hline
        ISSUE & The GitHub issue number or pull request sometimes is referenced in commit messages.\\\hline
        EMAIL & Contact e-mails of reviewers and authors usually appear after tags such as `Reported-by`.\\\hline
        URL & Vulnerability reports or blog posts; and bug-trackers references.\\\hline
        SHA & Sometimes commit hashes are mentioned to reference where the vulnerability was introduced.\\\hline
        VERSION & Software versions are sometimes important to find mentions of 
        malicious software.\\\hline
        SEVERITY & Vulnerability severity: low, medium, high, critical.\\\hline
        DETECTION & Vulnerabilities are detected manually or using specific tools (such as codeql, coverity, oss-fuzz, libfuzzer, and more).\\\hline
        SECWORD & Words or group of words that were identified as security-relevant 
        in previous work~\cite{10.1145/3463274.3463331,10.1145/3475716.3475781}.\\\hline
    \end{tabular}
    \label{tab:entities}
\end{table}

The extractor uses a pre-trained model provided by spacy to parse and
extract important features of the text. Some of these features, such as the 
part-of-speech tags, are used in the NER rules to improve precision. For
instance, an action implies a verb; therefore, we only extract keywords
like ``fix'', ``patch'', ``prevent'', etc, when the speech tag is a verb.
After extracting the different entities per section, we apply the different
sets of section rules. Rules are explained in more detail here: \url{https://tqrg.github.io/secomlint/#/secomlint-rules}. For some rules, 
the compliance checker takes into account the types of entities 
that are expected to obtain. For instance, for the rule \texttt{header-ends-with-vuln-id},
it is expected that the header line (first line of the message) ends
with an entity type of \texttt{VULNID}.

We also use the NER extractor to help the security engineer to
make the body section security informative, i.e., the tool 
provides the option \texttt{--is-body-informative} which 
checks if the body includes any security-related words. If not, 
it should be improved to be more clear. The body section is 
especially important because it's where security engineers 
explain the problem, its importance, and fix.

\subsection{Using SECOMlint}

\texttt{SECOMlint} can be either
executed by checking out its source code or---more
conveniently---be installed from the Python Package 
Index\footnote{PyPi: \url{https://pypi.org/}} via the \texttt{pip}
utility tool. The tool is meant to be used as a command-line 
application to ease its integration in the software development 
lifecycle. This section aims to provide an overview of the tool's 
functionalities. One can get an overview of 
all command-line arguments by using the \texttt{--help}
option after installing \texttt{SECOMlint}.

\begin{lstlisting}[label={lst:help},frame=tlrb]{Name}
$ secomlint --help 
\end{lstlisting}

Please note that to run \texttt{SECOMlint} 
successfully, the spacy English large model 
(\texttt{python -m spacy download 
en\_core\_web\_lg}) has to be downloaded since it 
is responsible for extracting text features, such as the part-of-speech tags that are later
used by the rule matcher. \texttt{SECOMlint} 
without any argument expects a text input. One way 
to provide it is by collecting the raw commit message with git: \texttt{git log -1 --pretty=\%B}. But if you want to run it on a simple string, you can also do it by using \texttt{echo}. 

\begin{lstlisting}[label={lst:help},frame=tlrb]{Name}
$ git log -1 --pretty=%B | secomlint
\end{lstlisting}

\begin{lstlisting}[label={lst:help},frame=tlrb]{Name}
$ echo "<message>" | secomlint
\end{lstlisting}

\texttt{SECOMlint} will output a report 
with the compliance of the message against 
each of the rules and a summary of it ``found X problem(s), Y warning(s);``. Problems are compliance violations that, in our view, need to be fixed, while warnings are smaller problems that should be fixed. 

\begin{figure}[t!]
    \centering
    \includegraphics[width=\linewidth]{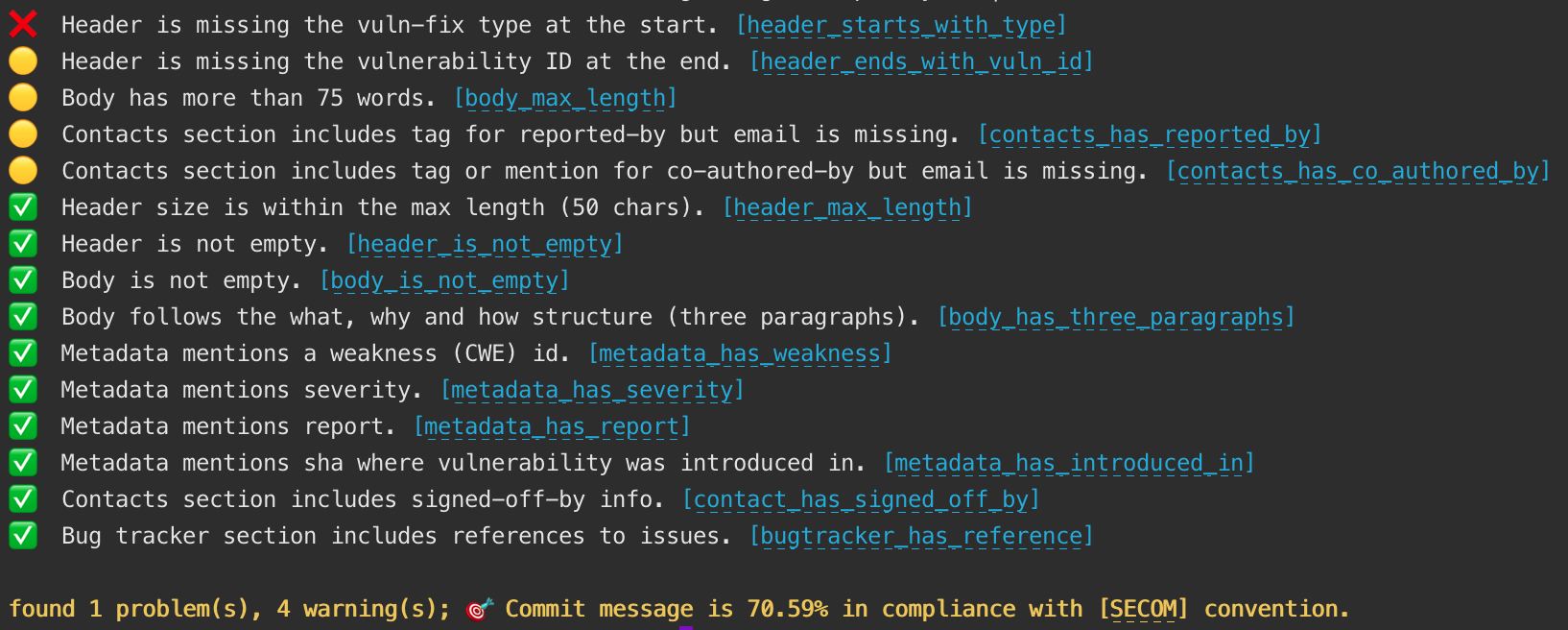}
    \caption{Example of the compliance report provided by \texttt{SECOMlint} (\texttt{--score} argument enabled).}\label{fig:output}
    \vspace{-1em}
\end{figure}

The importance of the different checkers can be customized with an \texttt{YAML} file. The \texttt{--config} argument should be used to pass the new configuration. By default, rules are all active. Therefore, if one wants to disable a rule, it will have to set the rule \texttt{active} tag to false. Problems are set as \texttt{type} equal to 1 and warnings as \texttt{type} equal to 0. Therefore, one can change the importance type to 1 if one wants to make a rule mandatory to comply with. The value checked for some rules can be customized with a value or regular expression. One example is the rule \texttt{header\_starts\_with\_type}, the value for the type at the beginning of the header can be changed.  

\begin{lstlisting}[label={lst:help},frame=tlrb]{Name}
header_starts_with_type:
  type: 1
  value: 'fix'
metadata_has_detection:
  active: false
\end{lstlisting}

\begin{lstlisting}[label={lst:help},frame=tlrb]{Name}
$ git log -1 --pretty=%B | secomlint \
                    --config=config.yml
\end{lstlisting}

Since the list of results may be a little bit extensive for some viewers, we added the argument \texttt{--no-compliance}, which will only show the result of the rules that do not comply with the convention, i.e., warnings and problems. In addition, to measure the compliance of his security commit message against the SECOM convention, one can use the argument \texttt{--score}. This argument will add the percentage of compliance to the summary (Figure~\ref{fig:output}). The score is calculated as a simple probability, i.e., the number of rules satisfied by the security commit message divided by the total number of rules that should be satisfied.

\begin{lstlisting}[label={lst:help},frame=tlrb]{Name}
$ git log -1 --pretty=%B | secomlint \
                    --no-compliance \
                    --score 
\end{lstlisting}

As mentioned before, the message's body is very 
important because it summarizes the problem,
its importance, and the patch. This information will 
help maintainers and researchers make better
understanding of the vulnerabilities and respective
solutions.
In our empirical analysis, we observed several cases where 
we couldn't understand how the commit was related 
to security because the vocabulary was not security-related enough. Therefore, we provide
an extra feature that runs the entity extraction on the body's message and checks if it can extract any security-related word. If not, then the writer should improve the body's message.

\begin{lstlisting}[label={lst:help},frame=tlrb]{Name}
$ git log -1 --pretty=%B | secomlint \
                     --is-body-informative
\end{lstlisting}

The tool can also be applied to a \texttt{.csv} file containing a column for messages by passing the file's name with the \texttt{--from-file} argument. 

\section{Evaluation}

We evaluated \texttt{SECOMlint} in two different
distributions: 1) \texttt{pre\_secom}, a subset of $500$ random security commit messages collect from 
NVD and OSV reports (excluding the automated ones); 2) \texttt{after\_secom}, a subset of $500$ random security commit messages collected from an experiment performed by a security engineer where several security vulnerability fixes used the SECOM convention to document the patch~\cite{blackhat}.
The mean compliance score for the \texttt{before\_secom} sample was $76.01$\%. The extractor extracted all types of entities from commit messages, but only a total of 2675 entities---an average of $5$ entities per security commit message.
The mean compliance score for the \texttt{after\_secom} sample was $86.61$\%. The tool extracted $10$ out of $12$ types of entities from commit messages, a total of $9911$ different entities---an average of $20$ entities per security commit message. It seems that by applying the SECOM convention, we can extract more information from security commit messages.

\section{Conclusions}

In this work, we have summarized the inner workings and features of \texttt{SECOMlint}. This tool was 
designed to be quickly introduced into the software 
development life cycle and support the appliance of the SECOM convention to security-related commit messages. By providing \texttt{SECOMlint} as open-source, we hope to foster the development of well-structured and more informative security commit 
messages. Further information on \texttt{SECOMlint}, its documentation, and source code are available through \url{https://tqrg.github.io/secomlint/}. A tool demonstration can be observed on YouTube: \url{https://youtu.be/-1hzpMN_uFI}. In the future, we plan to make the tool more customizable, extensible, and able to generate suggestions for the writer or even potential automated refactorings.

\bibliographystyle{IEEEtran}  
\bibliography{refs}  

\end{document}